\documentclass[aps,prl,superscriptaddress]{revtex4-1}
\usepackage{geometry}
\usepackage{setspace}
\usepackage{times}
\usepackage{fullpage}
\usepackage{color}
\usepackage{graphicx}
\begin{document}

\title{Hall-plot of the phase diagram for Ba(Fe$_{1-x}$Co$_x$)$_2$As$_2$}

\author{Kazumasa\,Iida}
\email[Electronical address:\,]{iida@nuap.nagoya-u.ac.jp}
\author{Vadim\,Grinenko}
\email[Electronical address:\,]{v.grinenko@ifw-dresden.de}
\affiliation{Department of Crystalline Materials Science Graduate School of Engineering, Nagoya University, Furo-cho, Chikusa-ku, Nagoya 464-8603, Japan}
\affiliation{IFW Dresden, P.O. Box 270116, 01171 Dresden, Germany}
\author{Fritz\,Kurth}
\affiliation{IFW Dresden, P.O. Box 270116, 01171 Dresden, Germany}
\affiliation{Dresden University of Technology, Faculty for Natural Science and Mathematics, 01062 Dresden, Germany}
\author{Ataru\,Ichinose}
\author{Ichiro\,Tsukada}
\affiliation{Central Research Institute of Electric Power Industry, 2-6-1 Nagasaka, Yokosuka, Kanagawa 240-0196, Japan}
\author{Eike\,Ahrens}
\affiliation{IFW Dresden, P.O. Box 270116, 01171 Dresden, Germany}
\affiliation{Dresden University of Technology, 01062 Dresden, Germany}
\author{Aurimas\,Pukenas}
\author{Paul\,Chekhonin}
\author{Werner\,Skrotzki}
\affiliation{Dresden University of Technology, 01062 Dresden, Germany}
\author{Angelika\,Teresiak}
\author{Ruben\,H\"{u}hne}
\affiliation{IFW Dresden, P.O. Box 270116, 01171 Dresden, Germany}
\author{Saicharan\,Aswartham}
\affiliation{IFW Dresden, P.O. Box 270116, 01171 Dresden, Germany}
\author{Sabine\,Wurmehl}
\affiliation{IFW Dresden, P.O. Box 270116, 01171 Dresden, Germany}
\affiliation{Dresden University of Technology, 01062 Dresden, Germany}
\author{Ingolf\,M\"{o}nch}
\affiliation{IFW Dresden, 01171 Dresden, Germany}
\author{Manuela\,Erbe}
\author{Jens H\"{a}nisch}
\affiliation{IFW Dresden, P.O. Box 270116, 01171 Dresden, Germany}
\affiliation{Karlsruhe Institute of Technology, Institute for Technical Physics, Hermann-von-Helmholtz-Platz 1, 76344 Eggenstein-Leopoldshafen, Germany}
\author{Bernhard\,Holzapfel}
\affiliation{Karlsruhe Institute of Technology, Institute for Technical Physics, Hermann-von-Helmholtz-Platz 1, 76344 Eggenstein-Leopoldshafen, Germany}
\author{Stefan-Ludwig\,Drechsler}
\author{Dmitri V. \,Efremov}
\affiliation{IFW Dresden, P.O. Box 270116, 01171 Dresden, Germany}

\date{\today}

\makeatletter

\maketitle

\section*{Abstract}
\noindent{\bf
The Hall effect is a powerful tool for investigating carrier type and density. For single-band materials, the Hall coefficient is traditionally expressed simply by $R_{\rm H}^{-1} = -en$, where $e$ is the charge of the carrier, and $n$ is the concentration. However, it is well known that in the critical region near a quantum phase transition, as it was demonstrated for cuprates and heavy fermions, the Hall coefficient exhibits strong temperature and doping dependencies, which can not be described by such a simple expression, and the interpretation of the Hall coefficient for Fe-based superconductors is also problematic. Here, we investigate thin films of Ba(Fe$_{1-x}$Co$_x$)$_2$As$_2$ with compressive and tensile in-plane strain in a wide range of Co doping. Such in-plane strain changes the band structure of the compounds, resulting in various shifts of the whole phase diagram as a function of Co doping. We show that the resultant phase diagrams for different strain states can be mapped onto a single phase diagram with the Hall number. This universal plot is attributed to the critical fluctuations in multiband systems near the antiferromagnetic transition, which may suggest a direct link between magnetic and superconducting properties in the BaFe$_2$As$_2$ system.}

\section*{Introduction}
It is widely believed that for most unconventional superconductors, Cooper pairs are mediated by spin or orbital fluctuations. A very good example is given by Co-doped BaFe$_2$As$_2$, one of the most studied Fe-based superconductors (FBS), in which the neutron resonance peak was observed\,\cite{Inosov}. Note that this resonance peak is hardly elucidated by electron-phonon interaction. Typically, the parent compound of Fe-based superconductor (FBS) shows a spin-density wave (SDW) phase at low temperatures. This magnetic instability is linked to the Fermi surface (FS) nesting between hole-like pockets centered at the $\Gamma$-point and electron-like pockets at M-points in the Brillouin zone\,\cite{Kordyuk2014}. Upon carrier doping, the nesting condition is deteriorated and the superconductivity appears at a given doping level. The emergence of superconductivity in the vicinity of SDW immediately pushed the idea of strong spin fluctuations providing the main glue for Cooper pairing in FBS.

External pressure\,\cite{Takahashi}, chemical pressure\,\cite{Kasahara01}, and strain in thin films\,\cite{Jan} may change the nesting conditions similarly to carrier doping. For the latter case, tensile or compressive in-plane strain with biaxial and uniaxial components is induced by the lattice and/or thermal expansion mismatch between film and substrate. Therefore, tensile or compressive in-plane strain may act as control parameter for the phase diagram by selecting a specific substrate.

Here, we report a systematic study of Ba(Fe$_{1-x}$Co$_x$)$_2$As$_2$ epitaxial thin films grown on MgO(001) and CaF$_2$(001) single crystalline substrates by pulsed laser deposition (PLD). The former substrate induces tensile strain, whereas the latter yields compressive one. Using transport data, we construct the phase diagram for Ba(Fe$_{1-x}$Co$_x$)$_2$As$_2$ thin films under different strain states (i.e., tensile and compressive in-plane strain). The resultant phase diagrams show that the N\'eel temperature ($T_{\rm N}$) and the superconducting transition temperature ($T_{\rm c}$) at a given Co doping level depend strongly on the direction of in-plane strain. Both $T_{\rm N}$ at zero doping and $T_{\rm c}$ at optimal doping level are enhanced by in-plane compressive strain in comparison to single crystals. Moreover, the whole phase diagram is shifted in the direction of higher Co doping. For tensile strain, $T_{\rm N}$ at zero doping is reduced and $T_{\rm c}$ at optimal doping level is almost unchanged, and the whole phase diagram is shifted to lower doping level. Finally, we demonstrate that the phase diagrams of all considered strained films and single crystals (i.e., the relaxed samples) including magnetic and superconducting regions can be mapped onto a single phase diagram with the Hall number as new variable. Our findings may suggest a direct link between magnetism and superconductivity in FBS.

\section*{Results}
\subsection*{Structural properties}
All Ba(Fe$_{1-x}$Co$_x$)$_2$As$_2$ films ($0\leq x\leq0.15$) were epitaxially grown on MgO(001) and CaF$_2$(001) substrates with high phase purity. The epitaxial relation is (001)[100]$_{\rm film}$$\|$(001)[100]$_{\rm MgO}$ and \sloppy(001)[110]$_{\rm film}$$\|$(001)[100]$_{\rm CaF_2}$. More information for the structural analyses by x-ray diffraction can be found in the Supplementary Information. As shown in Fig.\,\ref{fig:figure01}a, the lattice constant $a$ of the Ba(Fe$_{1-x}$Co$_x$)$_2$As$_2$ films on CaF$_2$ substrates (Ba-122/CaF$_2$) is shorter than that of the bulk samples\,\cite{Ni}, whereas the opposite relation holds for the Ba(Fe$_{1-x}$Co$_x$)$_2$As$_2$ films on MgO substrates (Ba-122/MgO). As expected, an elongation of the $c$-axis for Ba-122/CaF$_2$ and a shrinkage of the $c$-axis for Ba-122/MgO were observed due to the Poisson effect, as shown in Fig.\,\ref{fig:figure01}b (i.e., unit cell volume is roughly conserved). These structural changes are due to the biaxial strain with average uniaxial components over sample volume, $\epsilon_{xx}=\epsilon_{yy}$. Here, the average lattice deformations in the tetragonal phase are defined as $\epsilon_{xx}=(a_{\rm film}-a_{\rm PLD\,target})/a_{\rm PLD\,target}$ and $\epsilon_{zz}=(c_{\rm film}-c_{\rm PLD\,target})/c_{\rm PLD\,target}$ along the $a$- and $c$-axis. In this way, the biaxial in-plane tensile strain acts similarly to uniaxial pressure along the $c$-axis, while the in-plane compressive strain acts as negative pressure along the $c$-axis. The average lattice deformations at room temperature (RT) for Ba-122/MgO along the $a$- and $c$-axis are $\epsilon_{xx}=5.9\,\times10^{-3}$ and $\epsilon_{zz}=-4.4\,\times10^{-3}$, respectively. The corresponding values for Ba-122/CaF$_2$ films are $\epsilon_{xx}=-5.8\,\times10^{-3}$ and $\epsilon_{zz}=5.3\,\times10^{-3}$. It is noted that the lattice deformation for both films are almost constant irrespective of Co contents (Supplementary Fig.\,\ref{fig:figureS3}). The origin of the biaxial strain is discussed in the Supplementary Information.

Fig.\,\ref{fig:figure01}c summarizes the As position ($z$) in the unit cell for the strained films and PLD targets. The literature data for single crystals are also shown in the same figure\,\cite{Huang,Rotter,Rullier,Delaire}. It is apparent that the As coordinate is nearly independent on strain and Co doping.
 
\begin{figure}[h]
	\centering
		\includegraphics[width=13cm]{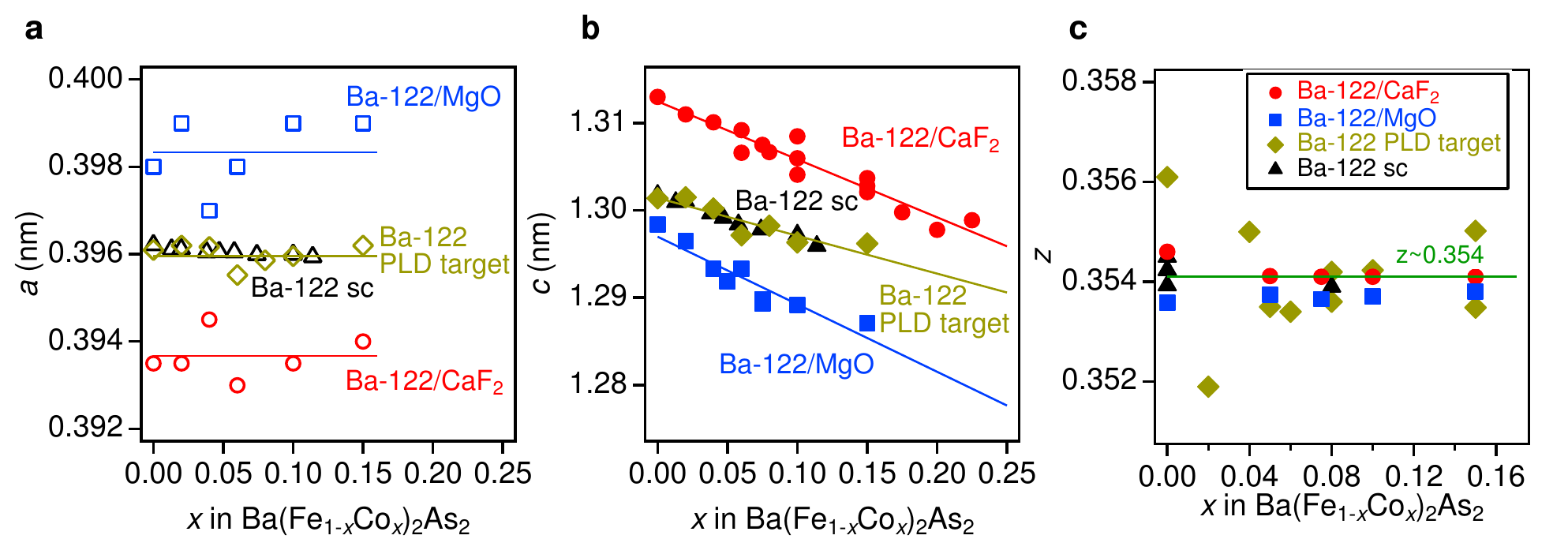}
		\caption{Co doping dependence of lattice parameters: (a) In-plane lattice constant $a$ of Ba(Fe$_{1-x}$Co$_x$)$_2$As$_2$ thin films on MgO and CaF$_2$ substrates, Ba(Fe$_{1-x}$Co$_x$)$_2$As$_2$ single crystals (Ba-122 sc)\,\cite{Ni}, and PLD target as a function of Co doping. The lines are a guide to the eye. 
(b) The corresponding out-of-plane lattice constants $c$ for the same samples. The lines are a guide to the eye. (c) The As position ($z$) for the strained films and PLD target materials as a function of the Co doping. The data of bulk single crystals are taken from Refs.\,\cite{Huang,Rotter,Rullier,Delaire}. The solid green line shows the average As position for unstrained samples.}
\label{fig:figure01}
\end{figure}
 
\subsection*{Resistivity and phase diagram}
\begin{figure}[b]
	\centering
		\includegraphics[width=8cm]{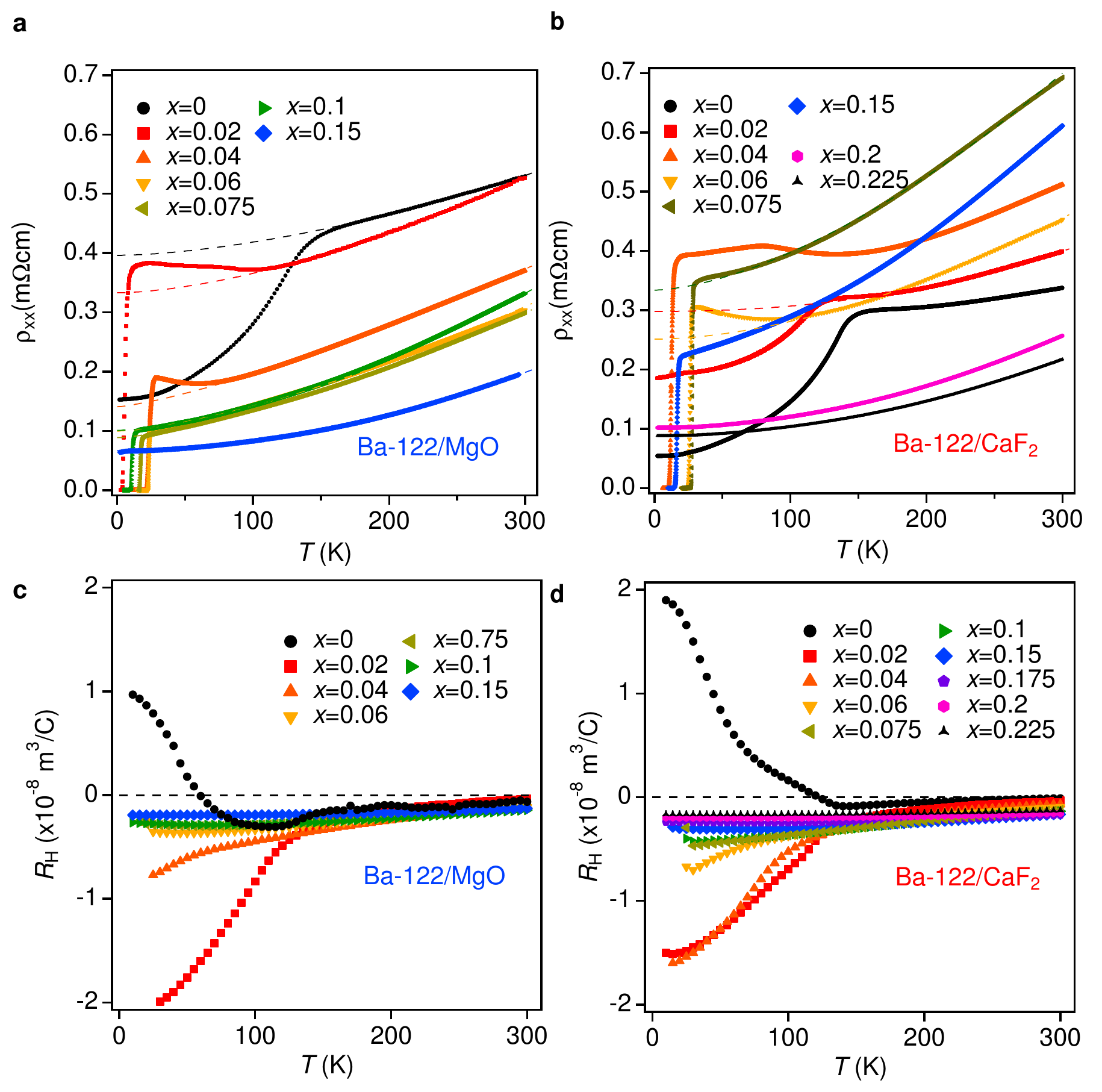}
		\caption{Transport properties of Ba-122/MgO and Ba-122/CaF$_2$ thin films: Resistivity data for Ba(Fe$_{1-x}$Co$_x$)$_2$As$_2$ thin films on (a) MgO and (b) CaF$_2$ substrates. Broken lines are the fitting curves using $\rho = \rho_0 + AT^n$ in the paramagnetic (PM) state. Hall coefficient of Ba(Fe$_{1-x}$Co$_x$)$_2$As$_2$ films on  (c) MgO and (d) CaF$_2$ substrates as a function of temperature.} 
\label{fig:figure2}
\end{figure}

The evolution of the in-plane longitudinal resistivity ($\rho_{xx}$) curves in zero magnetic field as a function of temperature for the Ba(Fe$_{1-x}$Co$_x$)$_2$As$_2$ films on MgO and CaF$_2$ substrates are displayed in Figs.\,\ref{fig:figure2}a and \ref{fig:figure2}b, respectively. The low-temperature state in the films on both substrates changes upon doping similar to the bulk material: from antiferromagnetic to superconducting, followed by metallic state\,\cite{Ni}. However, the doping levels at which the phase transitions occur depend on the strain state. For Ba-122/MgO, Co doping of $x=0.02$ induces superconductivity with a $T_{\rm c}$ of 7.5\,K. Additionally a sudden drop of the Hall coefficient around 100\,K due to the SDW transition was observed (see Fig.\,\ref{fig:figure2}c). Hence, for Ba-122/MgO with $x=0.02$ superconductivity coexists with antiferromagnetism. On the other hand, Ba-122/CaF$_2$ with the corresponding composition did not show superconductivity down to the lowest temperature available in our experiments (i.e., $\sim$ 2\,K).

\begin{figure}[b]
	\centering
			\includegraphics[width=8cm]{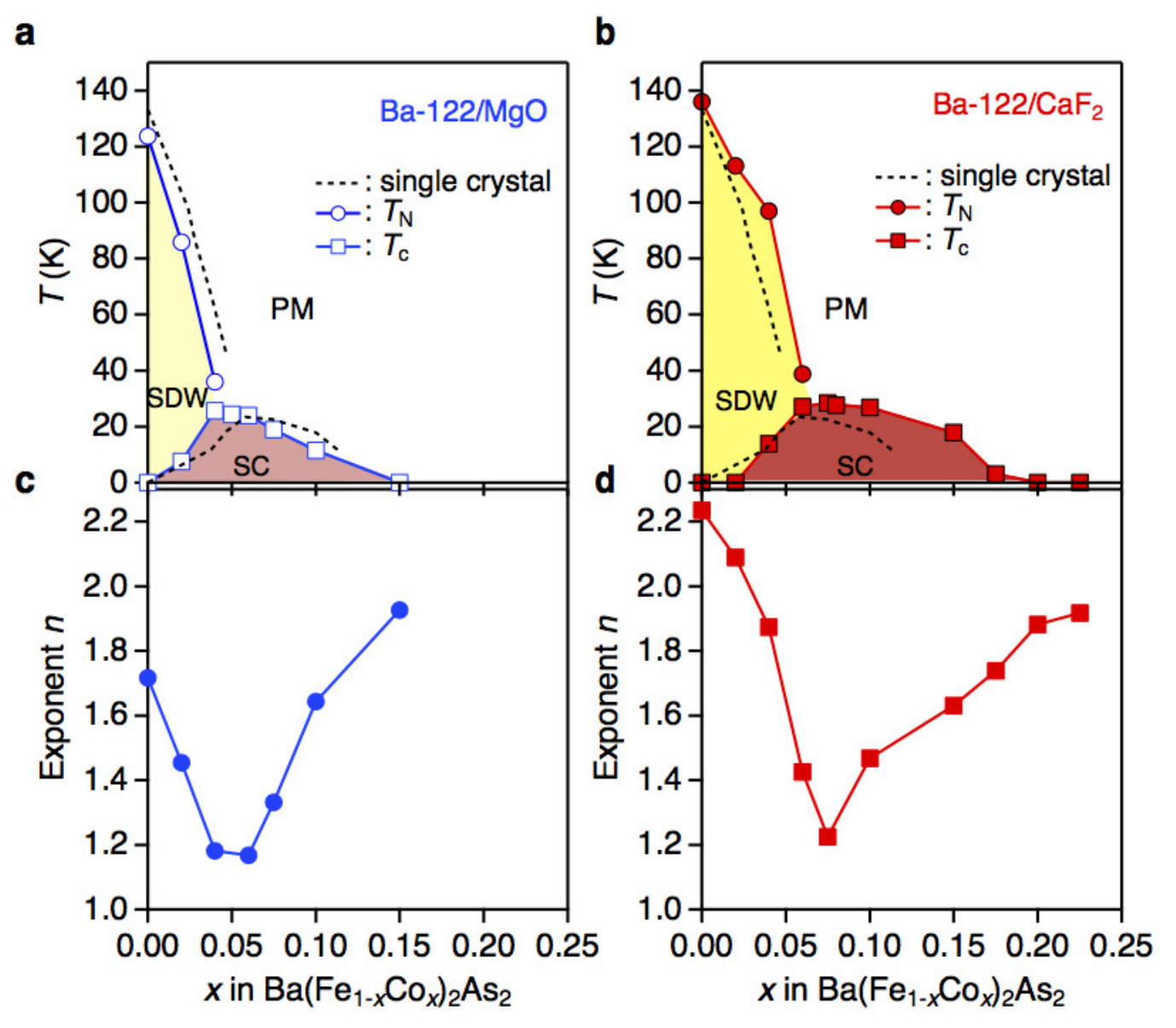}
		\caption{Electronic phase diagram of Ba(Fe$_{1-x}$Co$_x$)$_2$As$_2$: The electronic phase diagram of thin films grown on (a) MgO and (b) CaF$_2$ substrates. For comparison, the single crystal data\,\cite{Ni,Mun} are also shown in the figures as dotted lines. $T_{\rm N}$ and $T_{\rm c}$ denote the antiferromagnetic and the superconducting transition temperatures, respectively. SDW, PM, and SC are the spin density wave, paramagnetic, and superconducting phases, respectively. Value for the exponent $n$ taken from the resistivity data $\rho = \rho_0 + AT^n$ in the paramagnetic state: (c) Ba-122/MgO and (d) Ba-122/CaF$_2$.} 
\label{fig:figure3}
\end{figure}

Based on the resistivity data, the phase diagrams of both Ba-122/MgO and Ba-122/CaF$_2$ are constructed, Figs.\,\ref{fig:figure3}a and \ref{fig:figure3}b. For comparison, the single crystal data are plotted in the same figures. Here, $T_{\rm c}$ was determined by the superconducting onset temperature (Supplementary Fig.\,S7), whereas $T_{\rm N}$ was defined as a peak position of the temperature derivative of the resistivity curves in analogy to bulk single crystals (Supplementary Information in the section of criterion for $T_{\rm N}$)\,\cite{Pratt,Chu,Olariu}. It is noted that the peak position of the temperature derivative of the resistivity is related to the magnetic transition according to x-rays and neutron diffraction measurements\,\cite{Pratt}. Zero resistivity temperature and middle point of superconducting transition may be influenced by flux pinning effect. Therefore, we chose the onset temperature of resistivity as a criterion of the $T_{\rm c}$. It is clear from Fig.\,\ref{fig:figure3}a that tensile strain (Ba-122/MgO) slightly reduces $T_{\rm N}$ and shifts the superconducting dome to lower doping levels compared to the single crystals. A similar shift of the superconducting dome by in-plane tensile strain was observed in P-doped Ba-122 on MgO substrates\,\cite{Kawaguchi01}. To the contrary, biaxial in-plane compressive strain (Ba-122/CaF$_2$) effectively pushes the phase diagram to a higher doping level in comparison with single crystals (Fig.\,\ref{fig:figure3}b). Qualitatively, the shift of the phase diagram can be understood by examining the electronic band structure. At zero doping level, the {\it ab-initio} calculations show that compressive biaxial in-plane strain makes the band structure more two-dimensional with good nesting (see the section Discussion), resulting in a higher AFM transition temperature. Tensile strain shows the opposite effect which makes the band structure more three-dimensional, and consequently $T_{\rm N}$ decreases. For single crystals, a similar development of the FS takes place. Related angle-resolved photoemission spectroscopy (ARPES) measurements showed that upon Co doping the electronic states in the vicinity of the Fermi level become more three-dimensional\,\cite{Thirupathaiah}. Therefore, the two effects (charge doping and in-plane strain) determine the shift of the phase diagram along the doping axis.

The temperature dependence of the resistivity for the films in the paramagnetic (PM) state was fitted using $\rho = \rho_0 + AT^n$ and the resultant fitting curves are shown in Figs.\,\ref{fig:figure2}a and \ref{fig:figure2}b. This expression has been widely used for analyzing the resistivity in the quantum critical region, e.g.,\,\cite{Zhou,Shibauchi}.
The dependence of the power-law exponent $n$ on Co doping (i.e., $x$) is presented in Figs.\,\ref{fig:figure3}c and \ref{fig:figure3}d. For Ba-122/MgO, the exponent $n$ has a minimum value close to unity at $x\sim0.05$. This may be assigned to the AFM quantum critical point (QCP), where the AFM transition temperature goes to zero. For Ba-122/CaF$_2$, the QCP is observed at $x\sim0.075$ (Fig.\,\ref{fig:figure3}d). The presence of the AFM QCP for Co-doped Ba-122 has been proposed recently by specific heat, thermal expansion, and nuclear magnetic resonance measurements\,\cite{Ning,Meingast,Fernandes04}. The observed simultaneous shift of the QCP and the maximum $T_{\rm c}$ for the strained thin films may suggest the relationship between critical magnetic fluctuations and superconductivity in FBS.

\begin{figure}[h]
	\centering
			\includegraphics[width=6cm]{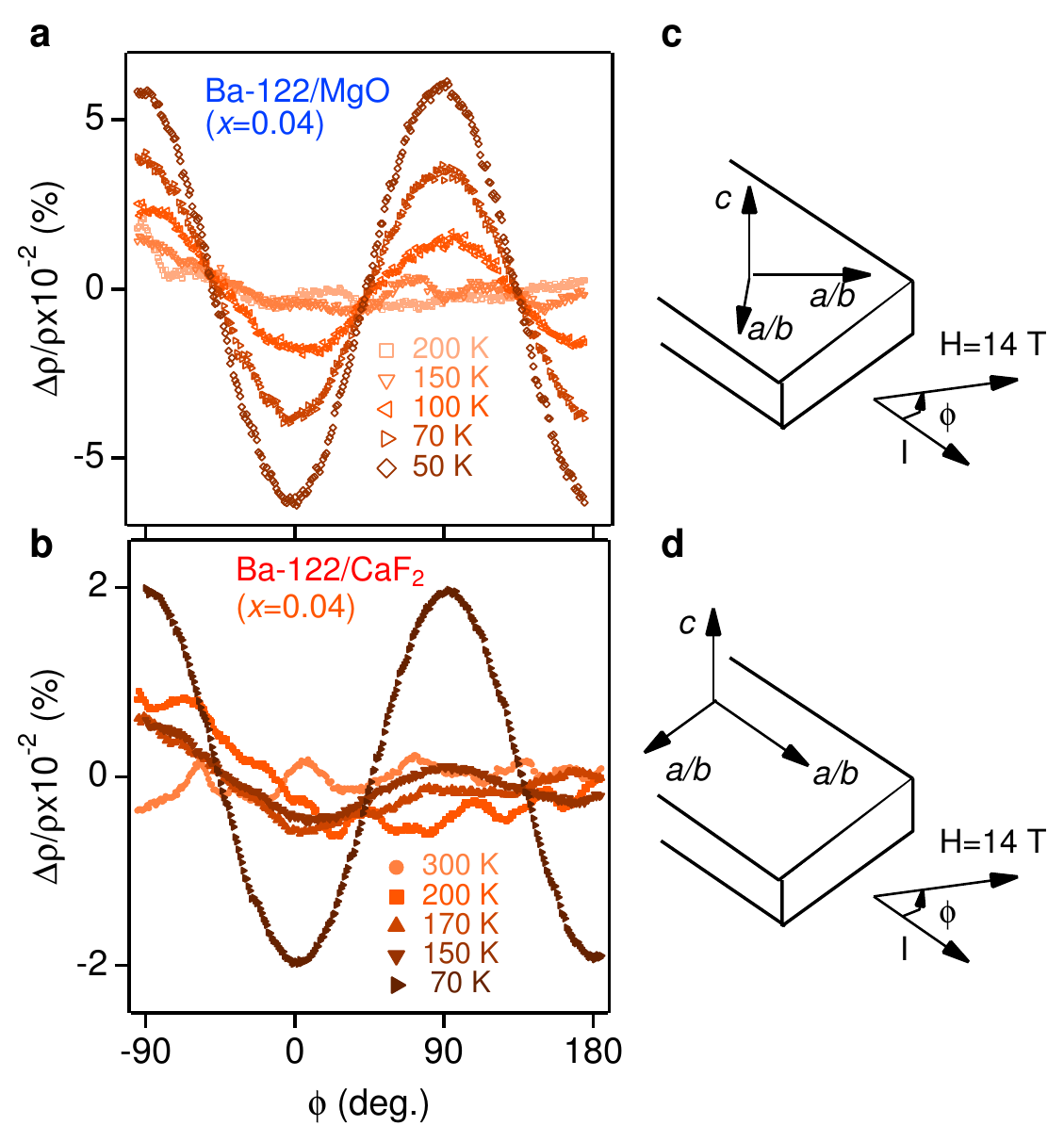}
		\caption{Angular dependence of in-plane magnetoresistance data: The angular dependence of in-plane magnetoresistance (MR) data ($\Delta \rho/\rho$) in the presence of a magnetic field (14\,T) for (a) Ba-122/MgO and (b) Ba-122/CaF$_2$. The sketch gives the orientation of the crystallographic axes for (c) Ba-122/MgO and (d) Ba-122/CaF$_2$ in orthorhombic notations.} 
\label{fig:figure4}
\end{figure}

In the case of thin films, the substrate may essentially weaken the orthorhombic distortion, as the Ba-122 grains are rigidly fixed at the interface by the substrate. This mechanism is responsible for the strain in the thin films (Supplementary Information). Another evidence for the tetragonal crystal structure being maintained in thin films grown on the substrates comes from the angular dependence of in-plane magnetoresistance (MR) measurements. An example of the in-plane MR for Ba-122/MgO and Ba-122/CaF$_2$ with the same Co doping level of $x=0.04$ in an applied field of 14\,T at various temperatures is shown in Figs.\,\ref{fig:figure4}a and \ref{fig:figure4}b. As stated above, Ba(Fe$_{1-x}$Co$_x$)$_2$As$_2$ thin films are grown on MgO(001) substrates with cube-on-cube configuration, whereas the basal plane of Ba(Fe$_{1-x}$Co$_x$)$_2$As$_2$ is rotated by 45\,$^\circ$ on CaF$_2$(001). We applied the current along the tetragonal [110] direction for Ba-122/CaF$_2$ and along the tetragonal [100] direction for Ba-122/MgO, respectively. According to Refs.\,\cite{Fisher,Chu-2}, a magnetic field of $B$ = 14\,T parallel to the $ab$-plane can partially detwin Ba(Fe$_{1-x}$Co$_x$)$_2$As$_2$ single crystals, leading to a twofold symmetry of the in-plane MR curves below the temperature at which the nematicity sets in. When bias current and magnetic field are parallel to the orthorhombic [100] or [010] axis, the in-plane MR curves show the maximum values. On the other hand, the position of the peak is shifted by 45$^\circ$, if the bias current ($I$) flows along orthorhombic [110] axis (zero angle corresponds to $B\|I$ ). In this case, the peak values are much smaller than for the former geometry (i.e., current and magnetic field $\|$ orthorhombic [100] or [010]). However, our results contradict the one obtained from single crystals. Below 100\,K, the measured in-plane MR curves for both films clearly follow an almost perfect sinusoidal angle dependence without phase shift. If the oscillation were defined by the nematic domains oriented by the applied field as in the case of single crystals\,\cite{Fisher,Chu-2}, the MR signal for Ba-122/MgO should be shifted by 45$^\circ$ with respect to the one for Ba-122/CaF$_2$. Additionally, the amplitude of MR signals for both films are quite small compared to those of single crystals. This indicates that the substrate completely blocks the rotation of the nematic/magnetic domains. However, the appearance of oscillation in the MR at a certain temperature $T^{\rm +}$ indicates some changes of the FS topology or scattering rates. The $T^{\rm +}$ is rather high compared to $T_{\rm N}$ and preserved at doping levels above the QCP of the SDW phase. By analogy with Refs.\cite{Kasahara02,Dhital2012}, $T^{\rm +}$ may be related to the nematic phase or fluctuating magnetic domains. This temperature is presumably increased by uniaxial strain if compared to relaxed Ba(Fe$_{1-x}$Co$_x$)$_2$As$_2$ single crystals.

\subsection*{Effective carrier density plot of the phase diagram}
The temperature dependencies of the Hall coefficients $(R_{\rm H})$ measured at 9\,T for Ba-122/MgO and Ba-122/CaF$_2$ films are shown in Figs.\,\ref{fig:figure2}c and \ref{fig:figure2}d. For both parent compound films (i.e., $x=0$), $R_{\rm H}$ is weakly decreasing with decreasing temperature until the SDW transition occurs, similarly to the observation in single crystals\,\cite{Fang,Mun}. In contrast to single crystals, however, $R_{\rm H}$ changes sign from negative to positive. This behavior can be understood qualitatively by the effect of strain on carrier mobilities. The non-doped Ba-122 is a compensated metal with equal electron and hole carrier densities. Therefore, a small change of the mobilities can strongly affect the experimental value of the effective $n_{\rm H}$ (especially in AFM state with reconstructed Fermi surfaces). Only a small amount of Co addition to the system leads to a drastic change in $R_{\rm H}$ at low temperature. For all films with $x=0.02$, $R_{\rm H}$ is decreased sharply with decreasing temperature below $T_{\rm N}$ due to a large change in the carrier concentration and mobility. This behavior is similar to that observed in single crystals\,\cite{Fang,Mun}.

\begin{figure}[b]
	\centering
    \includegraphics[width=8cm]{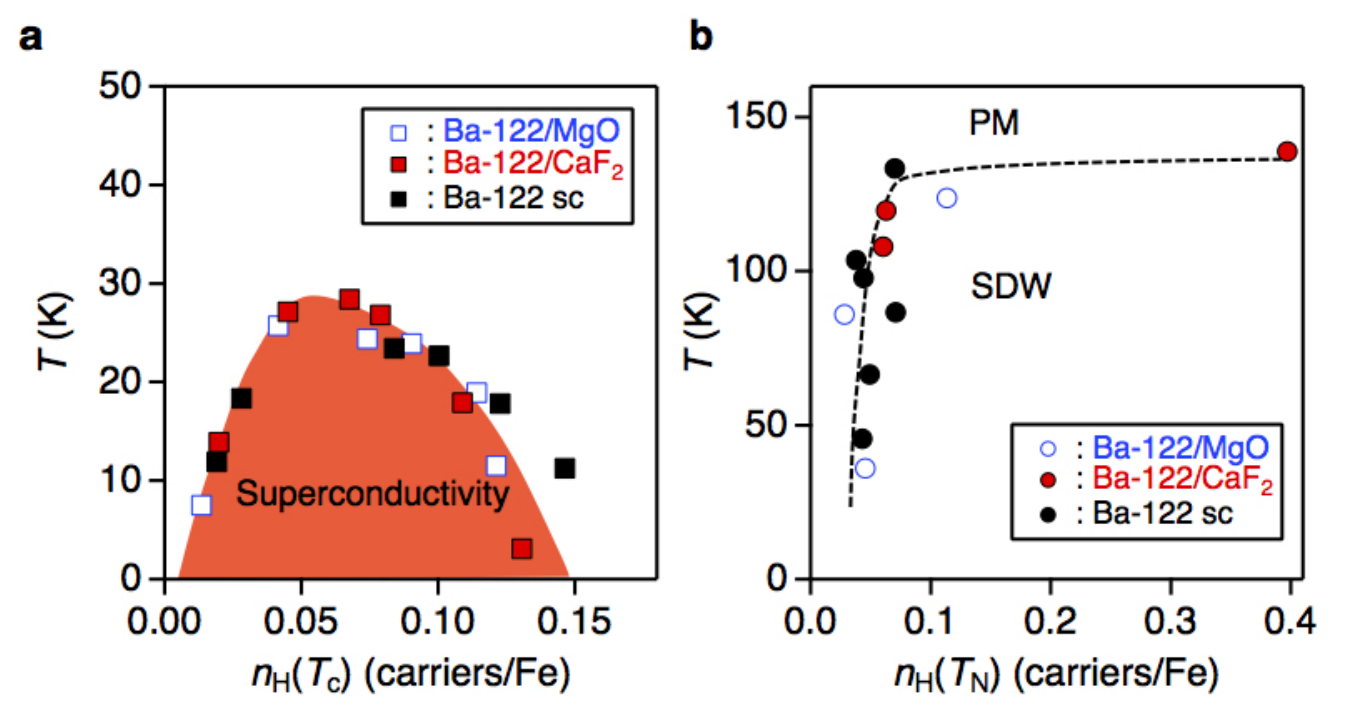}
		\caption{Effective carrier density plot of the superconducting $(T_{\rm c})$ and N\'eel $(T_{\rm N})$ temperatures of Ba(Fe$_{1-x}$Co$_x$)$_2$As$_2$: (a) Superconducting $(T_{\rm c})$ and (b) N\'eel $(T_{\rm N})$ temperatures as a function of $n_{\rm H}(T_{\rm c})$ and $n_{\rm H}(T_{\rm N})$. For comparison, Ba-122 single crystal data taken from Ref.\,\cite{Mun} are also plotted. SDW and PM are the spin density wave and paramagnetic phases, respectively. The labels show the characteristic range of $n_{\rm H}$ in SDW and PM phases.}
\label{fig:figure5}
\end{figure}

In order to quantify the effect of strain on the electronic properties, we consider the effective carrier number per Fe, $n_{\rm H}(T_{\rm c})$ and $n_{\rm H}(T_{\rm N})$, as $\frac{1}{e|R_{\rm H}|}\times \frac{V}{4}$, where $V$ is the unit cell volume estimated from Figs.\,\ref{fig:figure01}a and \ref{fig:figure01}b. Now, we re-plot $T_{\rm c}$ and $T_{\rm N}$ as a function of $n_{\rm H}(T_{\rm c})$ and $n_{\rm H}(T_{\rm N})$, as shown in Figs.\,\ref{fig:figure5}a and \ref{fig:figure5}b. For comparison, the single crystal data from Refs.\,\cite{Ni,Mun} are also shown in the graph. As can be seen, $T_{\rm c}$ for both strained thin films and single crystals can be mapped onto a master curve by $n_{\rm H}(T_{\rm c})$ as a new variable. Note that $n_{\rm H}$ scales both the position of the superconducting dome and the absolute value of $T_{\rm c}$, whereas the carriers numbers\,\cite{Ideta} and structural parameters (i.e., bonding angle and anion height)\,\cite{Johnston} scale only either the position of the superconducting dome or the absolute value of $T_{\rm c}$.

On the other hand, $T_{\rm N}$ is vaguely independent of $n_{\rm H}(T_{\rm N})$ for non-zero doping, i.e., a magnetic transition occurs, when $n_{\rm H}(T)$ approaches about 0.05 carriers/Fe.

\section*{Discussion}
The shift of the superconducting dome and the AFM transition temperature with uniform in-plane biaxial strain can be understood qualitatively by considering the effect of the strain on the FS shape, its orbital weight, and composition. First of all, our local (spin) density approximation (L(S)DA) calculation shows that the strain and the Co doping affect mainly the hole FS pockets located at the zone center, whereas the electron pockets at the zone corners are nearly unchanged. These results are consistent with the ARPES measurements for doped and undoped Ba-122 single crystals reported recently\,\cite{Neupane}. Additionally, it was observed that an L(S)DA and generalized gradient approximation (GGA) calculation gives a reasonable prediction of the effect of the Co doping on the band structure\,\cite{Neupane}. This observation strengthens our theoretical approach.

Analyzing the changes of the hole FS pockets, we found that $T_{\rm N}$ correlates well with the value of the $k_z$ dispersion of the Fe $3d$ $xz/yz$ orbitals on the hole FS pockets. As can be seen in Fig.\,\ref{fig:figure6}, the dispersion along $k_z$ of the undoped Ba-122 film on MgO increases compared to that of the undoped bulk sample (i.e., relaxed) due to tensile strain. $T_{\rm N}$ of the former is lower than that of the latter, as shown in Fig.\,\ref{fig:figure3}a. Simultaneously, the shape of the corresponding FS sheets is getting more three-dimensional. The same trend is observed for different Co doping but with fixed strain state. In contrast, compressive strain alone (i.e., Ba-122/CaF$_2$ with fixed Co doping) reduces the $k_z$ dispersion of the Fe $3d$ $xz/yz$ orbitals, which leads to the enhancement of $T_{\rm N}$. The observed three-dimensional effects of the FS are responsible for the suppression of the FS nesting conditions found by the ARPES study\,\cite{Neupane}. One can also see from Figs.\,\ref{fig:figure3}a and \ref{fig:figure3}b that the degree of shift in the superconducting dome along the doping axis correlates well with $T_{\rm N}$; lower/higher $T_{\rm N}$ pushes the SC dome towards the underdoped/overdoped region. Such a tendency may be understood if Cooper pairing and magnetic ordering is controlled by a common parameter.

The result of the band structure calculation is insufficient for the interpretation of the observed behavior shown in Fig.\,\ref{fig:figure5}b. The scaling indicates that the value of the effective carrier number $n_{\rm H}(T_{\rm N})$ at the phase transition is not only related to the electronic structure but also strongly affected by critical fluctuations which are not included in the band structure calculations. As was shown by Kontani $et$\,$al$., $n_{\rm H}$ scales with the antiferromagnetic (AF) correlation length ($\xi^{-2}$) in the case of strong AF spin fluctuations\,\cite{Kontani01}. Therefore, the value of $n_{\rm H}$ should approach zero at the phase transition. However, $n_{\rm H}(T_{\rm N})$ tends to a finite value $\sim0.05$ carriers/Fe (excluding films with zero doping level) at the magnetic transition as can be seen in Fig.\,\ref{fig:figure5}b. This seeming contradiction may be explained by considering the multiband nature of the FS. The Hall coefficient is defined as $R_{\rm H} = \sigma_{xy}/H(\sigma_{xx} \sigma_{yy})$, where $\sigma_{xy}$ and $\sigma_{xx (yy)}$ are the full conductivities summed up over all bands. Therefore, the Hall number is given by
$n_H = |\frac{(\sum_i\sigma_{h,i}+\sum_j\sigma_{e,j})^2}{\sum_i\sigma_{h,i}^{2}/n_{h,i}-\sum_j\sigma_{e,j}^{2}/n_{e,j}}|$,
where $\sigma_{h(e),i(j)}$ is the conductivity of a hole (h) or an electron (e) band, $i$ or $j$ is the summation index running over different hole or electron bands, respectively, and $n_{h(e)}$ is the corresponding carrier density\,\cite{Katayama}. Hence, as in the case of resistivity\,\cite{Grinenko2014}, the bands with the largest conductivities contribute most to the Hall number. At the phase transition, the conductivities of the interacting bands or part of the bands tend to zero due to strong scattering of their quasi-particles on the critical fluctuations. Therefore, $n_{\rm H}$ is simply determined by those parts of the FS which are less sensitive to critical fluctuations. This explains why the magnetic transition occurs at a finite $n_{\rm H}$ value. Good scaling for $T_{\rm c}$ is observed for each side of the superconducting dome, since those regions are far from the magnetic transition lines. In this case, the value of $n_{\rm H}(T_{\rm c})$ is sensitive to the distance to the SDW line. Therefore, the scaling for $T_{\rm c}$ can be interpreted by the indication of a strong interplay between $T_{\rm c}$ and the magnetic fluctuations slightly above $T_{\rm c}$, which are sensitive to the carrier doping and strain as well. A possible disorder effect on both $T_{\rm c}$ and $T_{\rm N}$ cannot be separated from the spin fluctuation effect, since impurity scattering is also included in $n_{\rm H}$.

In conclusion, we have shown that strain essentially affects the phase diagram of the generic system Ba(Fe$_{1-x}$Co$_x$)$_2$As$_2$. The biaxial in-plane strain is responsible for a nearly rigid shift of the whole phase diagram including the magnetic and superconducting regions along the electron doping. This behavior is explained by band structure calculations in which biaxial in-plane strain affects the FS similar to Co doping. Moreover, the superconducting dome is rigidly connected to the position of the SDW line. The direct relationship between the paramagnetic normal state and $T_{\rm N}$, as well as the relationship between $T_{\rm c}$ and the preceding state above $T_{\rm c}$ are given by the unusual plot of $T_{\rm N}$ and $T_{\rm c}$ with the Hall number at those temperatures. This emphasizes a crucial role of the critical fluctuations for superconductivity and magnetism in FBS. It is important to check whether a similar plot exists for other families of the FBS, too. Also, a microscopic explanation of the observed unusual behavior is still lacking. We believe that our experimental results will stimulate future theoretical investigations.

\section*{Methods}
\subsection*{Ba(Fe$_{1-x}$Co$_x$)$_2$As$_2$ films on MgO(001) and CaF$_2$(001) substrates}
Ba(Fe$_{1-x}$Co$_x$)$_2$As$_2$ films of around 100\,nm thickness have been grown on MgO(001) and CaF$_2$(001) substrates by pulsed laser deposition (PLD). PLD targets made by a solid state reaction with various Co levels ranging from $0\leq x\leq0.225$ were ablated by a KrF excimer laser with a laser repetition rate of 7\,Hz. Prior to the deposition, the substrates are heated to 850\,$^\circ$C. A base pressure of around 10$^{-8}$\,mbar at 850\,$^\circ$C is achieved, which increases to 10$^{-7}$\,mbar during the deposition. Thicknesses of all films ($\sim$ 100\,nm) were measured by scanning electron microscope images of cross-sectional focused ion beam (FIB) cuts. In a previous investigation on Ba(Fe$_{0.92}$Co$_{0.08}$)$_2$As$_2$ films (nominal composition $x=0.08$) by energy dispersive X-ray spectroscopy, we determined the Co content to 0.74$\pm$0.017, indicating good agreement with the nominal value\,\cite{Dario}.

\subsection*{Structural analyses by x-ray diffraction}
The $c$-axis texture and phase purity were investigated by x-ray diffraction in Bragg-Brentano geometry with Co-K$\alpha$ radiation. In-plane orientation of Ba-122/MgO and Ba-122/CaF$_2$ was investigated by using the 103 pole in a texture goniometer operating with Cu-K$\alpha$ radiation. In order to precisely evaluate the lattice parameter $a$ of Ba-122/MgO and Ba-122/CaF$_2$, high resolution reciprocal space maps (RSM) around the 109 reflection were performed with Cu-K$\alpha$ radiation. 

Temperature evolution of the lattice constants $c$ for Ba-122/MgO and Ba-122/CaF$_2$ was investigated by x-ray diffraction in Bragg-Brentano geometry with Cu-K$\alpha$ radiation in flowing He gas atmosphere. Diffraction patterns were acquired at elevated temperatures from 298\,K to 773\,K (Supplementary Fig.\,S4).

\subsection*{Determination of As position ($z$)}
The As position of the PLD target materials was refined by Rietveld analysis using powder x-ray data. For the thin films, $z$ was calculated by using the experimental lattice constants {\it a} and {\it c}, shown in Figs.\,\ref{fig:figure01}a and \ref{fig:figure01}b, and the optimized As position in the paramagnetic state, which is the same method as described in Ref.\,\cite{Grinenko2014}. 

\subsection*{Transmission electron microscopy (TEM)}
The samples for TEM analysis were prepared using a FIB (SMI3050MS2) by cutting and milling the identical films used for transport measurements. The microstructure near the interface of Ba-122/MgO with $x=0.06$ was analyzed using JEOL JEM-2100F [Supplementary Fig.\,\ref{fig:figureS5}(b)].

\subsection*{In-plane transport measurements}
Prior to the micro bridge fabrication, the temperature dependence of the resistance for all films was measured by a 4-probe method, in which small pins are aligned co-linear on the film surfaces. After the measurements, the films were photolithographically patterned and ion-beam-etched to fabricate a small bridge of 100\,$\mu$m width and 0.41\,mm length for transport measurements. No changes in transport properties after the micro bridge fabrication have been found. Longitudinal and transverse resistance were measured with four-probe configuration by a Quantum Design physical property measurement system (PPMS) up to 14\,T.

\subsection*{Theoretical analysis}
\begin{figure}[h]
	\centering
    \includegraphics[width=10cm]{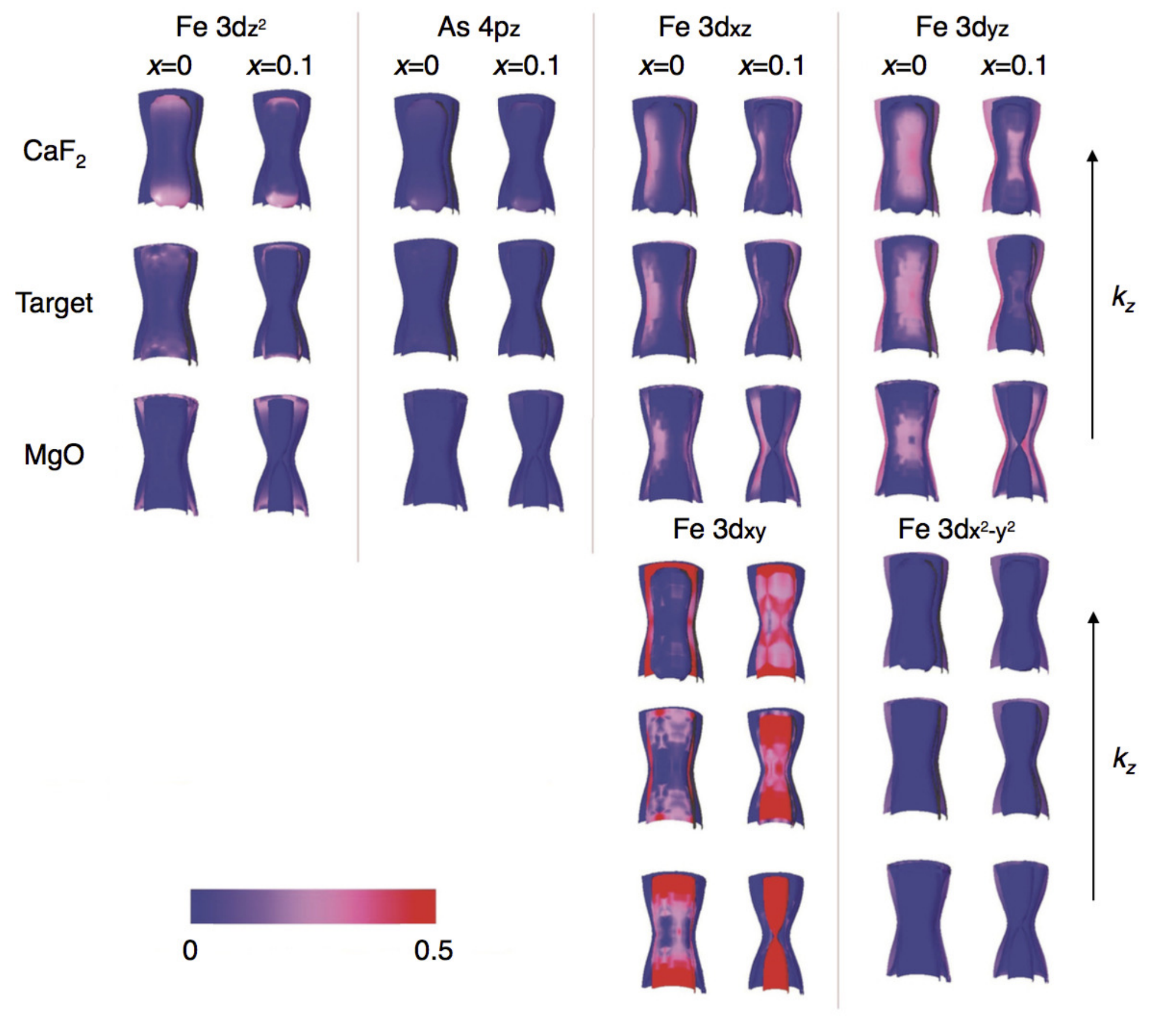}
		\caption{Fermi surface of Ba(Fe$_{1-x}$Co$_x$)$_2$As$_2$: Evolution of the Fermi surface (FS) of Ba(Fe$_{1-x}$Co$_x$)$_2$As$_2$ at the $\Gamma$ point as a function of Co doping and strain. The color code corresponds to a relative orbital weight per Fe-atom. The detailed theoretical approach can be found in Methods section.}
\label{fig:figure6}
\end{figure}

To understand the impact of the strain on the electron band structure, we performed density functional theory (DFT) calculations of the Fermi surface (FS) for both Ba(Fe$_{1-x}$Co$_x$)$_2$As$_2$ thin films and bulk single crystals. Our calculations were carried out within the local (spin) density approximation (L(S)DA) using the Full Potential Local Orbital band structure package (FPLO, http://www.fplo.de.)\,\cite{Koepernik}. The Co doping was taken into account within the virtual crystal approximation. As can be seen in Fig.\,\ref{fig:figure01}c, $z$ is almost constant irrespective of both strain and Co doping. Therefore, the absolute As position is just proportional to the lattice parameters. A $k$-mesh of $12\times12\times6$ $k$-points in the whole Brillouin zone was employed. The calculations were performed using the FPLO Ab-Initio Simulation Package within the Perdew, Burke and Ernzerhof (PBE)  functional for the exchange-correlation potential. The calculated hole FS for the Co doping level $x=0$ and 0.1 are summarized in Fig.\,\ref{fig:figure6}.

\subsection*{Acknowledgement} The authors thank Helge Rosner and Andrey Chubukov for fruitful discussions, Juliane\,Scheiter for preparing FIB cuts, as well as Michael\,K\"{u}hnel and Ulrike\,Besold for their technical support. The research has received funding from the European Union's Seventh Framework Programme (FP7/2007-2013) under grant agreement numbers 283141 (IRON-SEA) and number 283204 (SUPER-IRON). K.I. acknowledges support by JSPS Grant-in-Aid for Challenging Exploratory Research Grant Number 15K13336. A part of the work was supported by the DFG under the projects SPP1458 and GRK1621. S.\,W.\,acknowledges support by the DFG under the Emmy-Noether program (Grant no.\,WU595/3-1).

\subsection*{Authors contribution} K.I. and V.G. designed the study and wrote the manuscript together with D.V.E. and S.-L.D. The PLD targets were prepared by F.K., S.A., E.A. and S.W. Thin films were prepared by K.I. and F.K.\,~ K.I., F.K., A.T. and R.H. conducted x-ray experiments. A.P., P.C., and W.S. analyzed local strain of thin films by high-resolution EBSD. A.I. and I.T. conducted TEM investigation. M.E. and I.M. have developed micro bridge processing. K.I., F.K., V.G. and J.H. measured transport properties. D.V.E. and S.-L.D. developed a theoretical model and calculated the band structure. B.H., R.H., K.I., V.G., D.V.E., and S.-L.D. supervised the projects. All authors discussed the results and implications and commented on the manuscript at all stages.

\section*{Additional information}
The authors declare no competing financial interests. Correspondence and requests for materials should be addressed to K. I. and V. G.

\newpage

\clearpage
  
\begin{center}
\textbf{\large Supplemental Information}
\end{center}

\begin{center}
\noindent
\textbf{\large Hall-plot of the phase diagram for Ba(Fe$_{1-x}$Co$_x$)$_2$As$_2$}
\end{center}

\setcounter{equation}{0}
\setcounter{figure}{0}
\setcounter{table}{0}
\setcounter{page}{1}

\makeatletter
\renewcommand{\theequation}{S\arabic{equation}}
\renewcommand{\thetable}{S\arabic{table}}
\renewcommand{\thefigure}{S\arabic{figure}}
\renewcommand{\bibnumfmt}[1]{[S#1]}
\renewcommand{\citenumfont}[1]{S#1}
\maketitle

\subsection{Structural characterization by x-ray diffraction}
\begin{figure}[h]
	\centering
		\includegraphics[width=6.5cm]{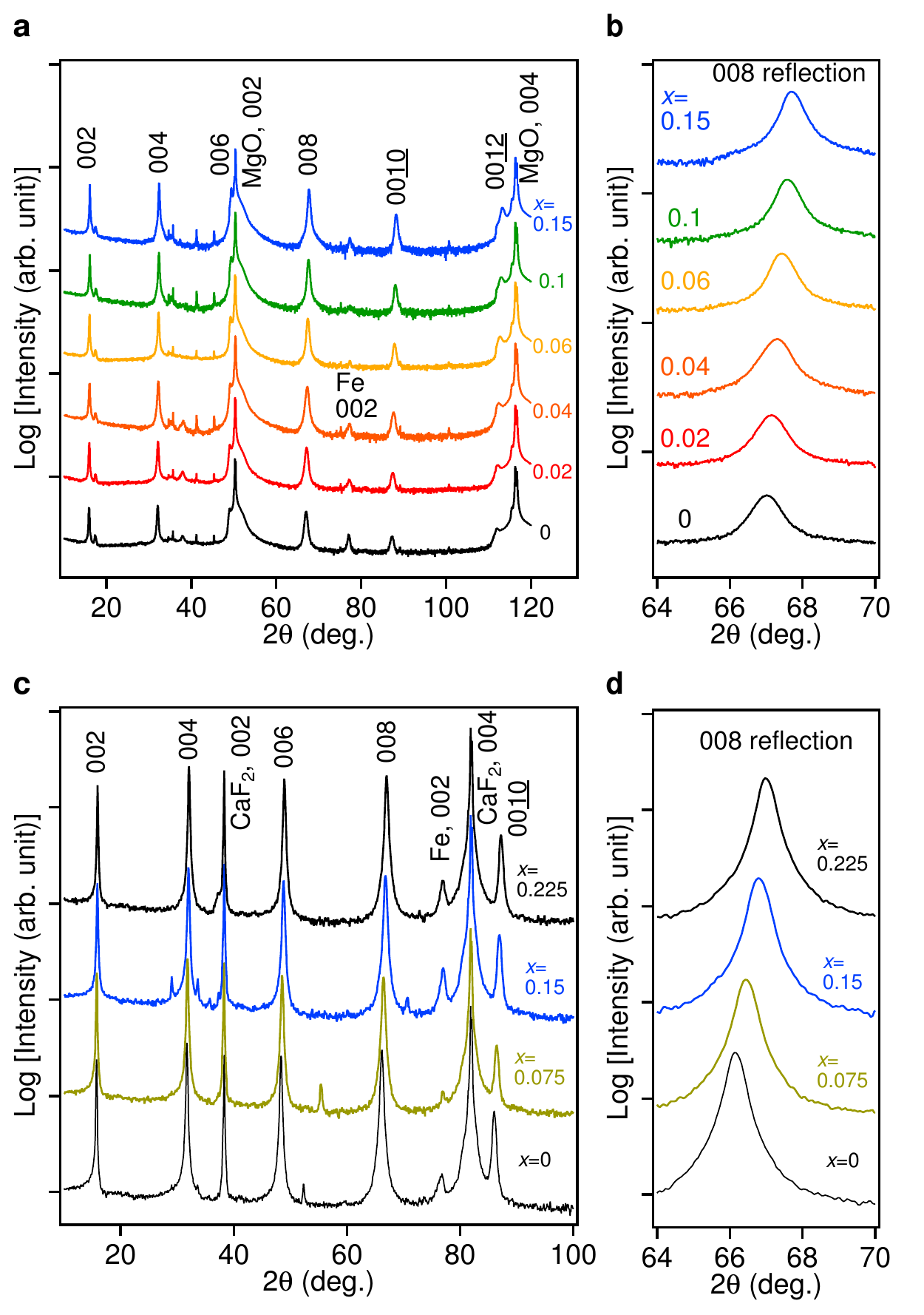}
		\caption{The $\theta-2\theta$ scan of the Ba(Fe$_{1-x}$Co$_x$)$_2$As$_2$ thin films on (a) MgO(001) and (c) CaF$_2$(001) substrates. The $\theta-2\theta$ scans in the vicinity of the 008 reflection for (b) MgO(001) and (d) CaF$_2$(001) substrates, respectively. A clear shift of the diffraction peak is observed with increasing Co content.}
\label{fig:figureS1}
\end{figure}

Our Ba(Fe$_{1-x}$Co$_x$)$_2$As$_2$ (Ba-122) thin films on MgO(001) and CaF$_2$(001) substrates have been grown by pulsed laser deposition (PLD). The composition of the films is almost identical to that of the Ba(Fe$_{1-x}$Co$_x$)$_2$As$_2$ PLD targets, indicative of a successful stoichiometric transfer\,\cite{Dario}. X-ray diffraction patterns for Ba(Fe$_{1-x}$Co$_x$)$_2$As$_2$ thin films on MgO and CaF$_2$ substrates as a function of the Co content are summarized in Fig.\,\ref{fig:figureS1}. Almost all peaks are assigned as 00$l$ reflections of Ba(Fe$_{1-x}$Co$_x$)$_2$As$_2$ and substrate, indicating a $c$-axis texture of Ba(Fe$_{1-x}$Co$_x$)$_2$As$_2$. For both cases, the 002 reflection of Fe is observed. The 008 peak position shifts towards higher angles with Co doping (Figs.\,\ref{fig:figureS1}b and \ref{fig:figureS1}d), as a result of a smaller lattice constant $c$ with increasing Co content.  

\begin{figure}[h]
	\centering
		\includegraphics[width=7.5cm]{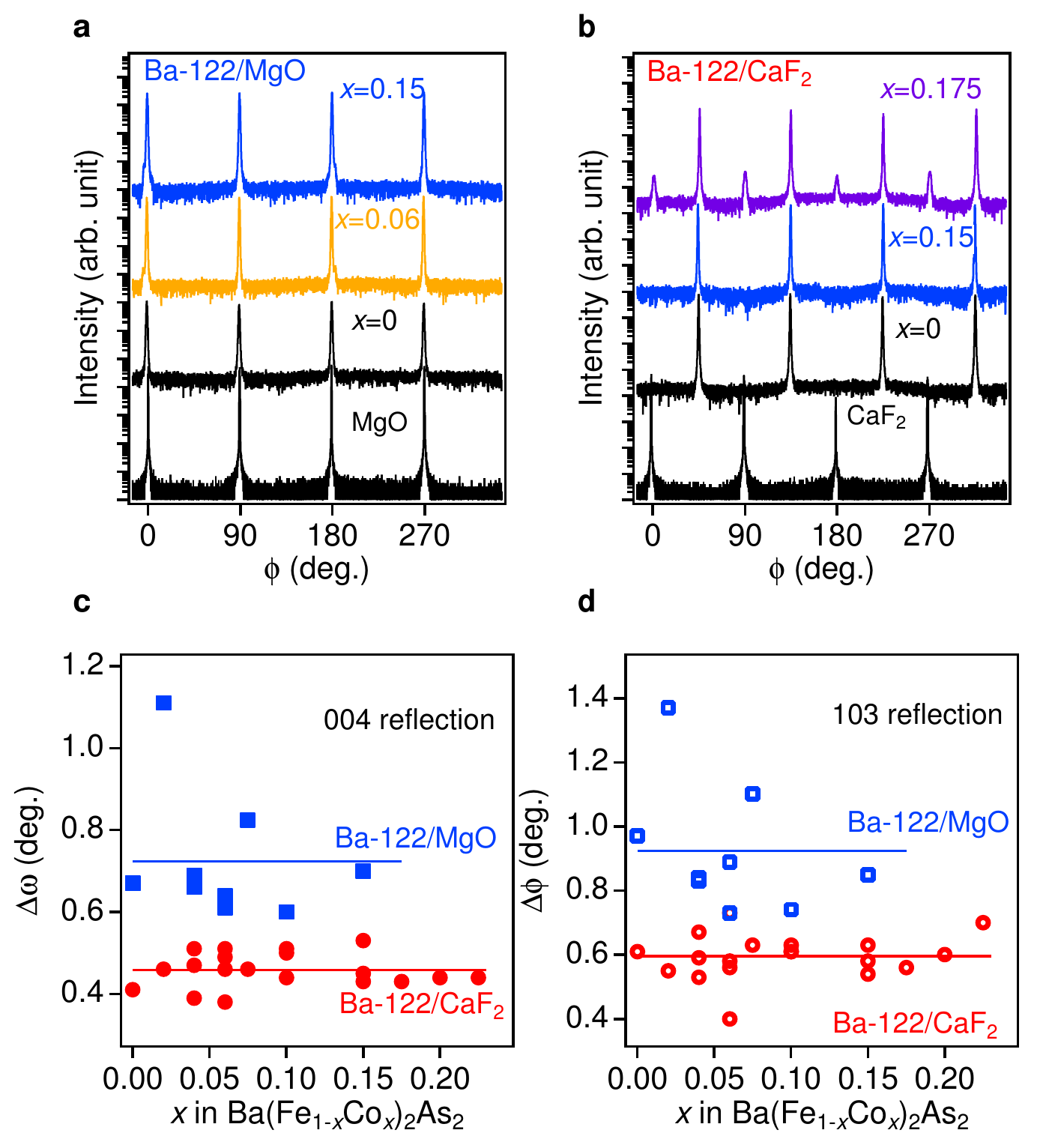}
		\caption{103 $\phi$-scans for Ba-122 thin films with different Co levels on (a) MgO and (b) CaF$_2$ single crystalline substrates. The 220 $\phi$-scans for MgO and CaF$_2$ substrates are also shown in the same graphs. (c) Full width at half maximum ($\Delta \omega$) value of the 004 reflection of Ba-122 thin films as a function Co content prepared on various substrates. (d) Average full width at half maximum ($\Delta \phi$) value of the 103 reflection of Ba-122 thin films as a function Co content prepared on various substrates.}
\label{fig:figureS2}
\end{figure}

$\phi$ scans of the 103 peak for Ba-122 films on both MgO and CaF$_2$ are summarized in Figs.\,\ref{fig:figureS2}a and \ref{fig:figureS2}b, respectively. For both films, $\phi$ scans of the 220 substrate reflection were also measured. Sharp and strong reflections from Ba-122 are observed at every 90$^\circ$. These results highlight that Ba(Fe$_{1-x}$Co$_x$)$_2$As$_2$ thin films are grown epitaxially. Here, the respective epitaxial relation for Ba-122 on MgO and CaF$_2$ substrates are (001)[100]$_{\rm film}$$\|$(001)[100]$_{\rm MgO}$ and (001)[110]$_{\rm film}$$\|$(001)[100]$_{\rm CaF_2}$.
The films on CaF$_2$ (Ba-122/CaF$_2$) with high doping regime ($x\geq0.175$) contained a small amount of in-plane 45\,$^\circ$ rotated domains. 

Ba-122/CaF$_2$ shows smaller full-width at half-maximum values of both out-of-plane and in-plane reflections compared to Ba-122/MgO, as shown in Figs.\,\ref{fig:figureS2}c and \ref{fig:figureS2}d. Here, the rocking curve of the 004 reflection was measured for all Ba-122 films. The $\Delta \omega$ and average $\Delta \phi$ values are mainly constant regardless of Co content.

\begin{figure}[b]
	\centering
		\includegraphics[width=10cm]{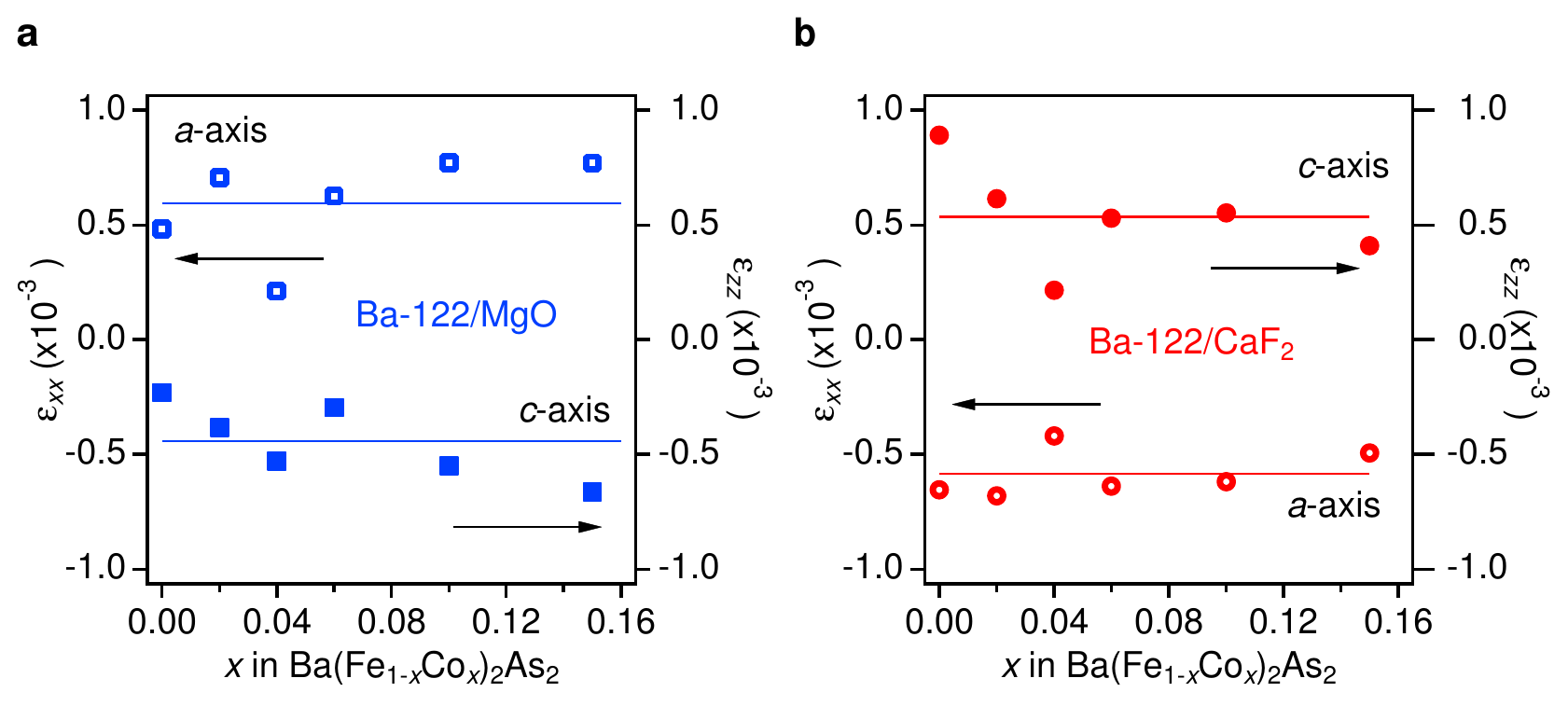}
		\caption{The lattice deformation in a tetragonal phase along the crystallographic $a$- and $c$-axis for Ba-122 thin films on (a) MgO and (b) CaF$_2$ substrates.}
\label{fig:figureS3}
\end{figure}

Lattice deformation in a tetragonal phase for both Ba-122/MgO and Ba-122/CaF$_2$ is summarized in Figs.\,\ref{fig:figureS3}a and \ref{fig:figureS3}b. Clearly, the lattice deformation for both films are almost constant  regardless of Co content.

\subsection{Origin of biaxial strain}
\begin{figure}[h]
	\centering
		\includegraphics[width=10cm]{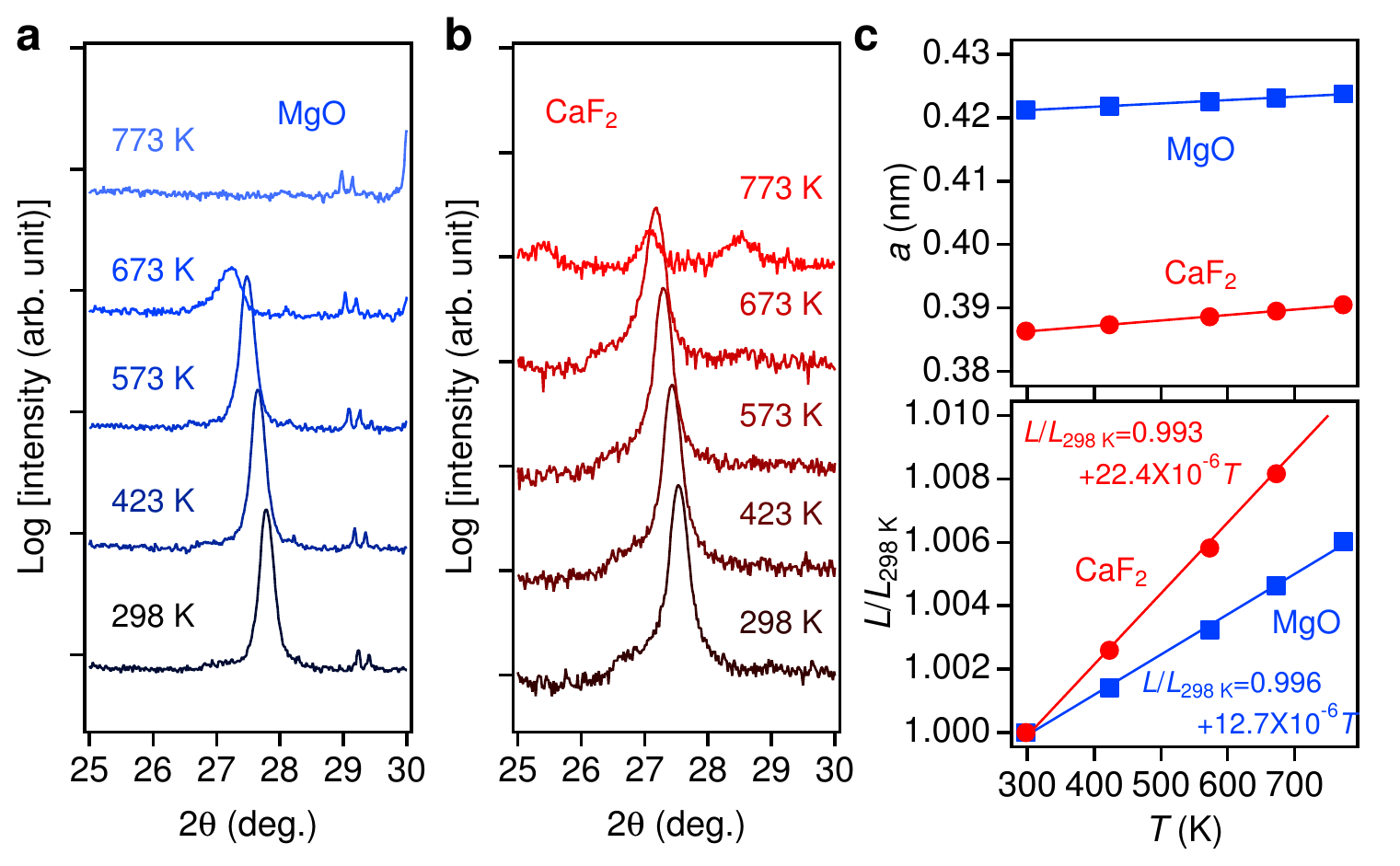}
		\caption{Temperature dependence of the x-ray diffraction patterns in the vicinity of the 004 reflection for (a) Ba-122/MgO and (b) Ba-122/CaF$_2$. (c) Temperature dependence of the lattice constants $a$ for MgO and CaF$_2$ substrates and the corresponding normalized values at 298\,K.}
\label{fig:figureS4}
\end{figure}

\begin{figure}[h]
	\centering
		\includegraphics[width=10cm]{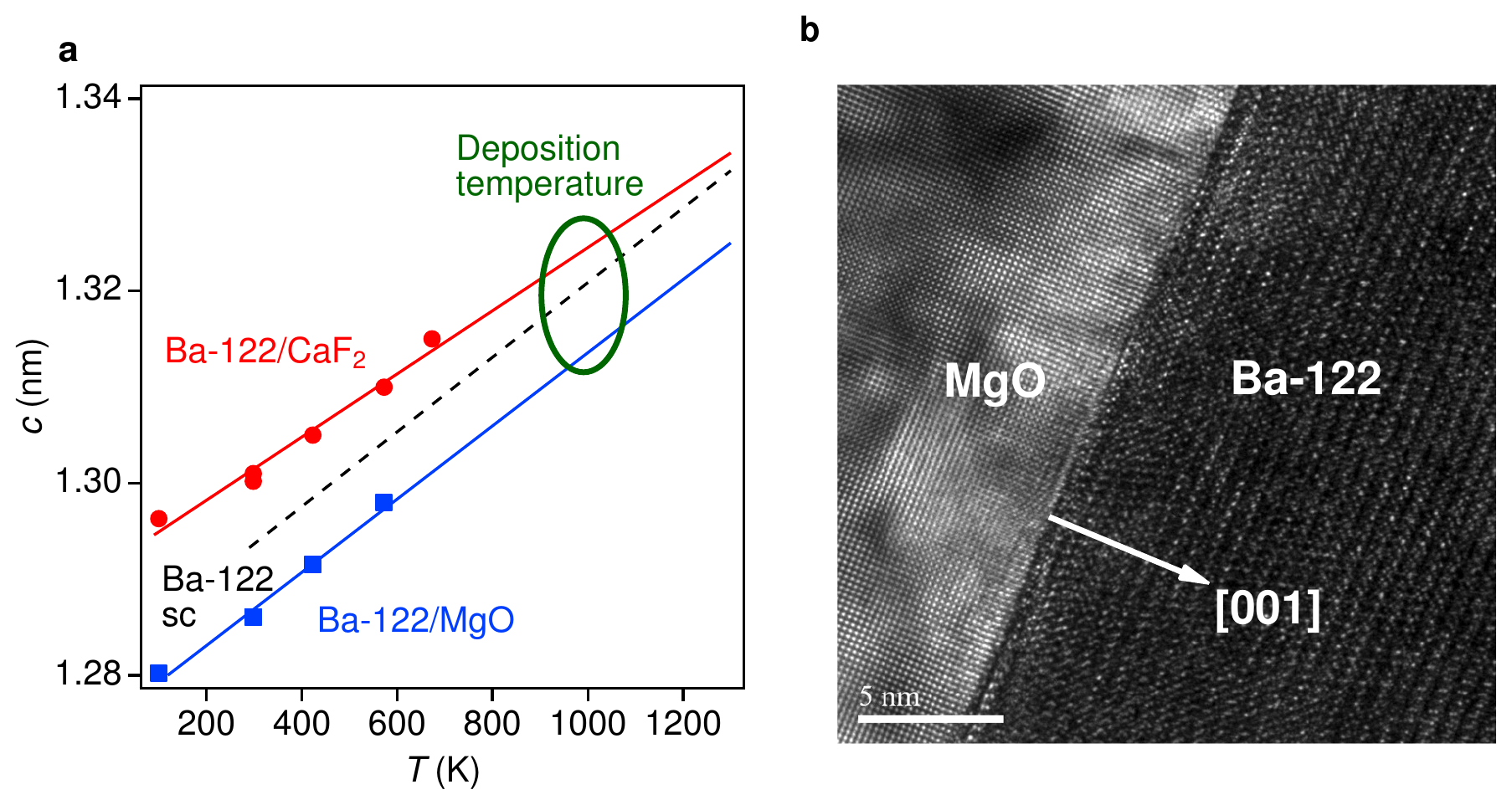}
		\caption{Temperature dependence of the out-of-plane lattice constants and microstructure: (a) Temperature dependence of the lattice constants $c$ for a Ba-122/MgO and a Ba-122/CaF$_2$ film with $x=0.15$. The single crystal data (dotted lines) were estimated from Ref.\,\cite{Luz02}. (b) TEM picture in the vicinity of the interface between the Co-doped Ba-122 film ($x=0.06$) and the MgO substrate.}
\label{fig:figureS5}
\end{figure}

The origin of the biaxial strain in thin films may be understood by considering the evolution of the lattice constants $c$ with temperature. The temperature dependence of x-ray diffraction patterns in the vicinity of the 004 reflection of Co-doped Ba-122 films with the same Co doping, $x=0.15$, but grown on different substrates MgO and CaF$_2$, are presented in Figs.\,\ref{fig:figureS4}a and \ref{fig:figureS4}b. Our measurements have been conducted in flowing He gas. For both Ba-122 films, the diffraction peaks are shifted toward lower angle, indicative of the elongation of the lattice constant $c$ due to thermal expansion. For Ba-122/MgO, the diffraction intensity is observed to decrease around 673\,K and almost disappeared at 773\,K, indicating that the Ba-122 phase was decomposed. On the other hand, for Ba-122/CaF$_2$ films the diffraction peak was still observed even at 773\,K, although the peak height is significantly reduced. From the temperature dependent x-ray measurements, the lattice parameters $a$ of MgO and CaF$_2$ substrates have also been evaluated, as shown in Fig.\,\ref{fig:figureS4}c. The evaluated values of the linear thermal expansion coefficients, $\alpha$, are given in Table\,\ref{tab:table1}. Based on those results, the temperature dependencies of the lattice constants $c$ for the two films are presented in Fig.\,\ref{fig:figureS5}a. For comparison, the temperature dependence of the $c$-axis length of a Ba-122 single crystal is also shown. These data were calculated using the experimental lattice constant at room temperature and the thermal expansion coefficient for the single crystal with $x=0.115$\,\cite{Luz02}. The lattice constants $c$ for the Ba-122/CaF$_2$ and the single crystal are close to each other at high temperature, indicating that the respective in-plane lattice parameters $a$ are also close to each other at that temperature. Upon cooling, the difference in the $c$-axis length between the Ba-122/CaF$_2$ and the single crystal increases, which is attributed to a large difference in the thermal expansion coefficients of the CaF$_2$ substrate and Ba-122 (Table\,\ref{tab:table1}). This effect is mainly responsible for the compressive strain of $\epsilon_{xx}=-5.8\,\times10^{-3}$ in Ba-122/CaF$_2$  thin films.

\begin{table}
\caption{\label{tab:table1}The linear thermal expansion coefficient, $\alpha$, of the MgO and CaF$_2$ substrates at 298\,K. The value for  Ba(Fe$_{0.885}$Co$_{0.115}$)$_2$As$_2$ along the crystallographic $a$-axis at 300\,K was taken from Ref.\,\onlinecite{Luz02}.}
\begin{ruledtabular}
\begin{tabular}{lccr}
 & MgO & CaF$_2$ & Ba(Fe$_{0.885}$Co$_{0.115}$)$_2$As$_2$ \\
\hline
$\alpha\rm\,(\times10^{-6}\,K^{-1})$ & $12.4$ &  $22.4$ & $8.5$ \\
\end{tabular}
\end{ruledtabular}
\end{table}

On the other hand, the two traces of the Ba-122/MgO film and the single crystal are almost parallel, presumably due to a small thermal expansion mismatch (Table\,\ref{tab:table1}). Unlike Ba-122/CaF$_2$ films\,\cite{Fritz01,Ichinose}, a clean interface between the Co-doped Ba-122 film and the MgO substrate has been observed by transmission electron microscopy, as shown in Fig.\,\ref{fig:figureS2}b and reported in Ref.\,\cite{Hiramatsu}. In this case, the lattice misfit yields a tensile strain in the thin films. However, the measured magnitude of the strain, $\epsilon_{xx}=5.9\,\times10^{-3}$, is smaller than expected for the relatively large lattice misfit of around -6\,\% at room temperature. This larger difference is caused by strain relaxation since our Ba-122 films have a thickness of about 100\,nm, which is beyond the critical thickness for relaxation. Indeed, our previous investigations on the Ba-122/Fe bilayer system revealed a critical thickness of around 30\,nm for a lattice misfit of -2.5\%\,\cite{Jan}, whereas the corresponding value of Ba-122/MgO results in a few atomic layers\,\cite{Hiramatsu}. Therefore, the presence of a small amount of biaxial strain, $\epsilon_{xx}=\epsilon_{yy}=5.9\,\times10^{-3}$, indicates that residual strain exists beyond the critical thickness, which was also observed in III-V semiconductors\,\cite{Westwood} and P-doped Ba-122 films on MgO substrates\,\cite{Kawaguchi01,Sakagami}. Additionally, we have found nanoscale oscillation of uniaxial strain components using high resolution electron backscatter diffraction (HR-EBSD)\,\cite{Wilkinson}, which will be discussed later. This strain state can be described by the modulation of $\epsilon_{xx}$ and $\epsilon_{yy}$ having opposite sign in the range of about $\pm0.2$\% with $\epsilon_{zz}\sim0$. We assume that this strain component originates from the formation of a low-angle grain boundary network during the coalescence of slightly rotated nanoscale islands nucleating on the mismatched MgO surface during film growth. These strain inhomogeneities were detected by HR-EBSD using a comparable experimental and evaluation procedure provided in reference\,\cite{Paul}.

\subsection{The local strain by HR-EBSD}
\begin{figure}[h]
	\centering
    \includegraphics[width=6cm]{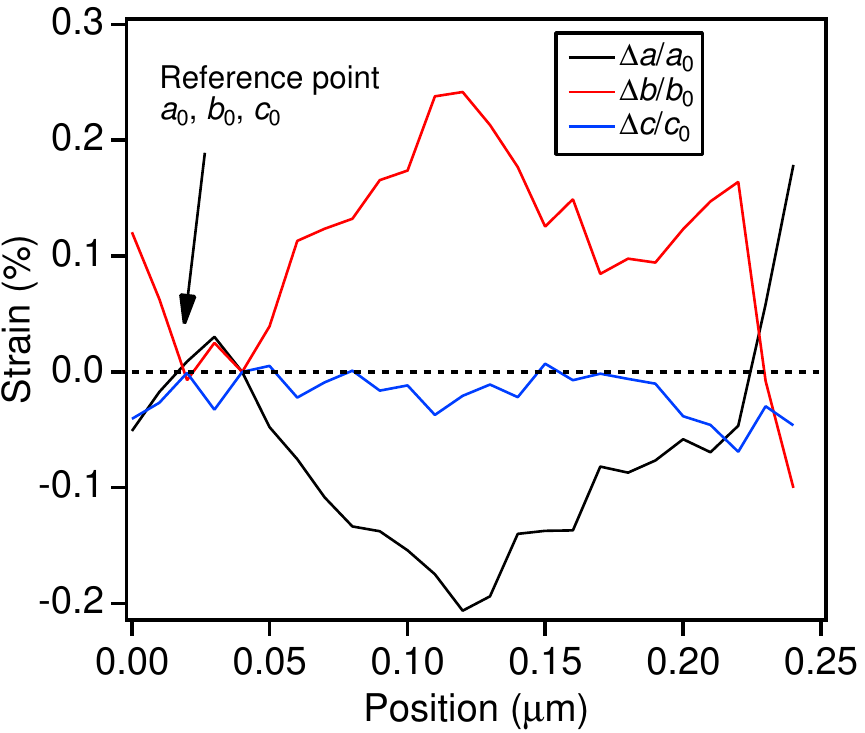}
		\caption{The local strain distribution of the crystallographic $a$, $b$, and $c$ direction as a function of position.}
\label{fig:figureS6}
\end{figure}

Local strain components in a Co-doped Ba-122 ($x=0.06$) thin film on MgO were measured by high resolution electron backscatter diffraction (HR-EBSD). HR-EBSD was performed as line scans in a Zeiss Ultra 55 scanning electron microscope using 20\,kV acceleration voltage and 10\,nm step size. The EBSD patterns have been recorded and analyzed subsequently with an in-house written software, based on the algorithm developed by Wilkinson $et$ $al$\,\cite{Wilkinson}. Shown in Fig.\,\ref{fig:figureS6} is the local strain distribution of the crystallographic $a$, $b$, and $c$ direction (normal strains) relative to a chosen reference position as a function of position. Variations of strain for all directions are within $\pm0.2$\%.
 
\subsection{Determining the superconducting transition temperature, $T_{\rm c}$}
\begin{figure}[h]
	\centering
    \includegraphics[width=5cm]{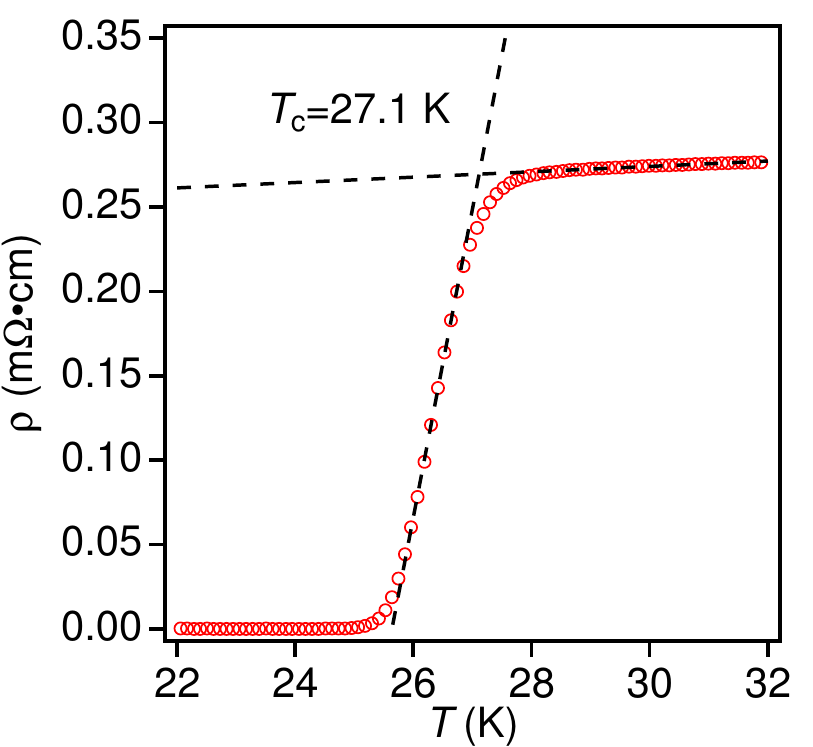}
		\caption{The resistivity curve of a 10\,\% Co-doped Ba-122 thin film on a CaF$_2$ substrate. $T_{\rm c}$ was determined to 27.1\,K.}
\label{fig:figureS7}
\end{figure}

The superconducting onset transition temperature was defined as the intersection between the linear fit of the normal state resistance and the steepest slope of the superconducting transition (see Fig.\,\ref{fig:figureS7}). Zero resistivity temperature and middle point of superconducting transition are influenced by flux pinning effect. Therefore, we chose the onset temperature of resistivity as a criterion of the $T_{\rm c}$. Our optimally Co-doped Ba-122 superconducting films showed an exact match of the zero resistivity temperature $T_{\rm c,0}$ and the onset $T_{\rm c}$ from magnetization measurements, proving high quality of our films\,\cite{Dario}.

\subsection{Determining the magnetic transition temperature, $T_{\rm N}$}
\begin{figure}[h]
	\centering
    \includegraphics[width=9cm]{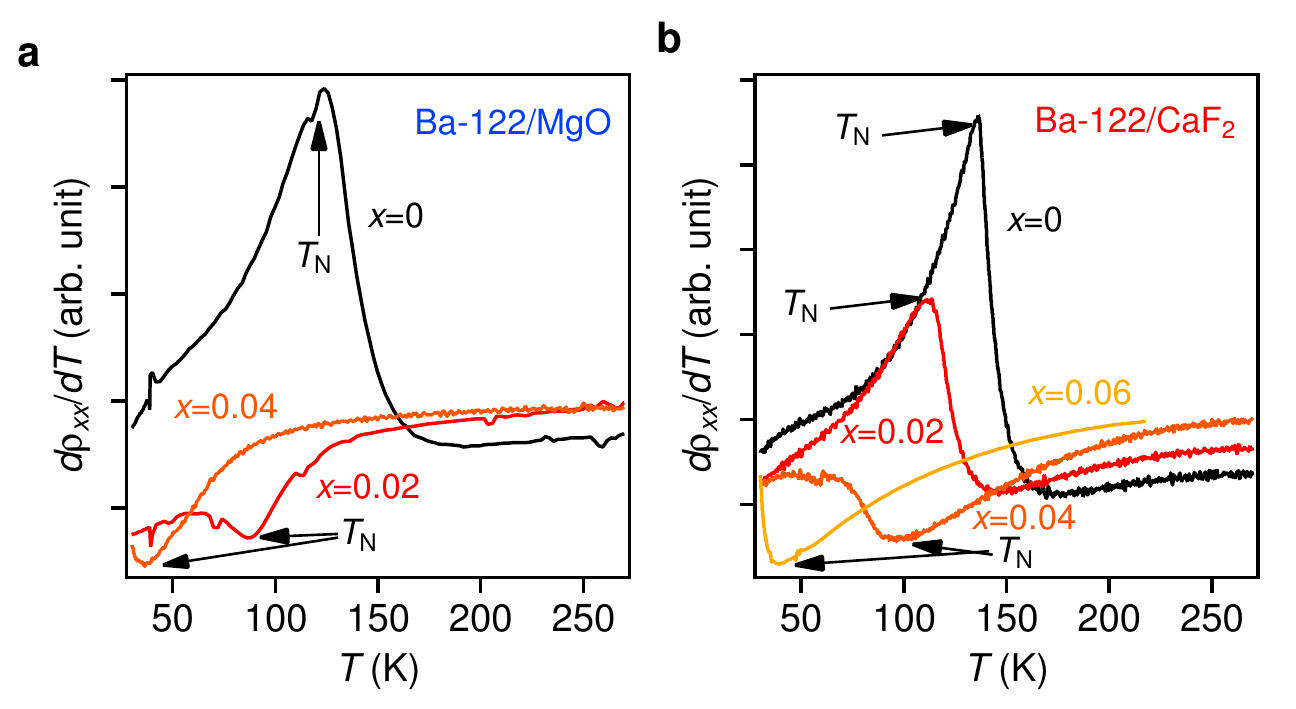}
		\caption{The temperature derivative of the resistivity curves of (a) Ba-122/MgO and (b) Ba-122/CaF$_2$. The peak position is assigned as $T_{\rm N}$}.
\label{fig:figureS8}
\end{figure}

The peak position of the temperature derivative of the resistivity is related to the magnetic transition according to x-rays and neutron diffraction measurements\,\cite{Pratt}. Therefore, the magnetic transition temperature, $T_{\rm N}$, was defined as the peak position, as shown in Fig.\,\ref{fig:figureS8}. In contrast to bulk single crystals (i.e., unstrained material), the kink above $T_{\rm N}$ related to the structural/nematic transition is absent. Recently, a similar behavior was observed under application of uniaxial strain to the $ab$-plane of Co-doped Ba-122 single crystals\,\cite{Fisher}. Therefore, the uniaxial component in the films obscures the nematic/structural transition (Fig.\,\ref{fig:figureS6}). However, inhomogeneous strain does not affect $T_{\rm N}$ and $T_{\rm c}$ noticeably, since the transitions are rather sharp in the strained films.

\end{document}